\documentclass{osa-article}
\usepackage{subcaption}
\usepackage{graphicx}
\usepackage[normalem]{ulem}
\usepackage{todonotes}
\journal{oe}

\articletype{Research Article}

\usepackage{lineno}

\begin{document}

\title{Physics-assisted Generative Adversarial Network for X-Ray Tomography}

\author{Zhen Guo\authormark{1},
Jung Ki Song\authormark{2},
George Barbastathis\authormark{2,3,7},\\
Michael E. Glinsky\authormark{4},
Courtenay T. Vaughan\authormark{4},
Kurt W. Larson\authormark{4}, 
Bradley K. Alpert\authormark{5},
and
Zachary H. Levine\authormark{6,8}\\}

\address{\authormark{1}Department of Electrical Engineering and Computer Science,
Massachusetts Institute of Technology,
Cambridge, Massachusetts, 02139,
USA\\
\authormark{2}Department of Mechanical Engineering,
Massachusetts Institute of Technology,
Cambridge, Massachusetts, 02139,
USA\\
\authormark{3}Singapore-MIT Alliance for Research and Technology (SMART) Centre, Singapore 138602\\
\authormark{4}Sandia National Laboratories,
Albuquerque, New Mexico, 87123, USA\\
\authormark{5}Applied and Computational Mathematics Division,
National Institute of Standards and Technology,
Boulder, Colorado, 80305, USA\\
\authormark{6} Quantum Measurement Division,
National Institute of Standards and Technology,
Gaithersburg, Maryland 20899, USA\\
\authormark{7} gbarb@mit.edu\\
\authormark{8} zlevine@nist.gov}



\begin{abstract}
X-ray tomography is capable of imaging the interior of objects in three dimensions non-invasively, with applications in biomedical imaging, materials science, electronic inspection, and other fields. The reconstruction process can be an ill-conditioned inverse problem, requiring regularization to obtain satisfactory results.
%
Recently, deep learning has been adopted for tomographic reconstruction. Unlike iterative algorithms which require a distribution that is known \textit{a priori}, deep reconstruction networks can learn a prior distribution through sampling the training distributions. In this work, we develop a Physics-assisted Generative Adversarial Network (PGAN), a two-step algorithm for tomographic reconstruction.
In contrast to previous efforts, our PGAN utilizes maximum-likelihood estimates derived from the measurements to regularize the reconstruction with both known physics and the learned prior. 
Compared with methods with less physics assisting in training, PGAN can reduce the photon requirement with limited projection angles to achieve a given error rate.
The advantages of using a physics-assisted learned prior in X-ray tomography may further enable low-photon nanoscale imaging.
\end{abstract}

\section{Introduction}
\label{intro}

X-ray tomography is a powerful method for imaging the internal details of objects in three dimensions non-invasively~\cite{wang2009metrology, bord2002x, mahmood2015real}, and it has wide applications in biomedical imaging~\cite{momose1996phase}, materials science~\cite{salvo20103d}, electronic inspection~\cite{alam2017impact}, and other fields.  The penetrating ability of X-rays makes it possible to obtain a series of two-dimensional Radon transforms (commonly known as radiographs) of the object viewed from different angles~\cite{withers2021x}. After capturing the radiograph measurements, objects can be reconstructed using a three-dimensional computed tomography algorithm.

The reconstruction process of X-ray tomography is generally an ill-conditioned inverse problem. This is because measurements taken at a discrete number of angles can only sparsely sample the high frequencies of the object. Therefore, full-angle measurements with a high sampling rate are preferred to best resolve the ambiguity in the inverse solution. However, in practice, limited-angle measurements are often used due to physical constraints, e.g., with laminar samples,
leaving entire sectors of the Fourier space unsampled~\cite{araya2018deep, doi:10.1137/0143028}.
For objects or samples that are radiation-sensitive, a low number of photons per scan is also preferred to minimize the total exposure and potential damage, making the effect from ill-conditioning even more severe.
In such cases, an analytic reconstruction algorithm such as Filtered Back Projection (FBP) is inadequate as it can generate reconstructions with noise and streak artifacts~\cite{frikel2013reconstructions}. Iterative algorithms whose objective function includes a term representing prior knowledge about the object may compensate for the deficits in Fourier space coverage and often produce higher fidelity results~\cite{Bouman1993}. This is understood as  regularizing the problem, i.e., reducing the space of possible solutions of the inverse problem to a subdomain in which the object must belong~\cite{sato1981tomographic,verhoeven1993limited}. 

When prior knowledge is used in an iterative algorithm, the optimization balances minimization of the residual of the simulated measurements from a reconstructed object against minimization of the regularization term.
Assumed priors such as 
sparsity, total variation, and nonlocal similarity priors have been used extensively in X-ray tomography~\cite{allag2018x, kazantsev2015employing, zhang2020multi}. However, without trial and error, it is not straightforward to choose the appropriate prior and regularization weight for a given set of objects~\cite{wang2020deep}. A prior distribution may also be learned from the dataset itself by a machine learning algorithm. Using a large amount of paired training data, a prior can be determined through exploring the statistical properties 
of the training distributions, improving the reconstruction quality. Recently, learned priors have been successfully applied to tomography in treating the ill-conditioned inverse problem. In particular, deep learning, a subset of machine learning that is based on artificial neural networks, achieved promising results~\cite{kobler2018variational, araya2018deep, wang2020deep, huang2020limited, bubba2019learning, huang2019data}. For example, efforts have been made in using learned priors from deep neural networks to recover boundary information~\cite{bubba2019learning}, and to generate missing projections with a data-consistent reconstruction method~\cite{huang2019data}. However, reports have shown that these methods suffer from reconstruction artifacts and instabilities~\cite{antun2020instabilities, gottschling2020troublesome}. To avoid these issues, some works leverage reconstructions from an analytic or iterative algorithm~\cite{wang2020deep, huang2020limited, schwab2019deep, goy2019high, kang2021dynamical}, or use a two-step deep learning strategy to generate reconstructions that are empirically more stable and accurate~\cite{wu2020stabilizing}.

In this paper, we develop a Physics-assisted Generative Adversarial Network (PGAN) for limited-angle X-ray tomography, and demonstrate the ability of the learned prior in imaging the structure of 3D integrated circuits at low-photon conditions. In contrast to the previous efforts, our PGAN utilizes a maximum-likelihood estimate (MLE) with known physics to compensate for the inherent ill-conditioning of the problem, especially the prevalence of shot noise of Poisson statistics in the low-photon measurements. The imaging geometry of the X-ray tomography and the Poisson statistics in low-photon measurements are considered in the first step of the reconstruction.
Then, a generative model uses the learned prior to further improve the maximum-likelihood estimate. Therefore, the PGAN reconstruction incorporates the known physics from an iterative algorithm 
and the learned prior from a deep generative model, drastically improving the reconstruction quality.
Note that our physics-assisted approach is different from the physics-informed approach where the physical laws are incorporated in the objective function~\cite{karniadakis2021physics, doi:10.1137/20M1354210, doi:10.1137/18M1225409}. Here, we follow a recent thread of research that leverages conventional algorithms to generate an input to the neural network which already satisfies the known physical constraints~\cite{goy2018low, Guo:22, kang2021dynamical, matlock2021physical}, empirically finding results with improved quality by separating the physics from the network optimization. By comparing with methods using no physics-assisting and less physics-assisting training strategies, we demonstrate that further separation of the physical priors may improve the effectiveness of the learned prior from a deep generative model.

To evaluate our reconstruction method, we propose a model dubbed CircuitFaker to produce synthetic circuits that are capable of emulating real-world integrated circuits (IC) with design rules. The implicit correlations of the circuits constitute the prior to be learned by the reconstruction algorithms. We simulate X-ray imaging using projections with Beer's Law attenuation. Then, we formulate four different variants of deep generative models, using the maximum-likelihood estimate from an iterative algorithm as the input approximant to include the known Poisson statistics in the measurements as well as the imaging geometry of the system. The output of the generative models is the reconstruction that has been regularized by the learned prior. We show that the learned prior from the deep generative models improves the reconstruction quality compared to methods with no physics-assisting and less physics-assisting if the photon flux is limited. The strategy without physics assisting is End-to-End training, where the generative model directly transforms the X-ray projections to the imaging object. The strategy with less physics assisting is using Filtered Back Projection as the input approximant to only include the known imaging geometry without knowing the Poisson statistics in the measurements. Training with the most physics assisting leads to high quality reconstruction which requires minimum number of photons per ray for a given error rate. This is the key result of our paper, suggesting that separation of the physical priors may improve the effectiveness of the learned prior from a deep generative model. Another key result of our paper is that CircuitFaker gives a prior distribution to be learned by our PGAN which reduces the search space of tomographic reconstruction. This is confirmed by the study of Bernoulli processes, where the objects are not spatially correlated. Thus, one does not need to have as strong illumination to obtain high quality reconstruction compared with maximum-likelihood estimation.

Our paper is structured as follows: first, we formulate the forward model for X-ray tomography in Section~\ref{forward-model}. Next, in Section~\ref{algo} we introduce the inverse algorithms. We explain the design of our PGAN in Section~\ref{PGAN}. We lay out our evaluation methods in Section~\ref{eval}. Finally, we present numerical results in Section~\ref{numerical}. Conclusion is in Section~\ref{conclusion}. A portion of this work was reported previously~\cite{guo2021advantage}.

\section{Forward model for X-ray tomography}
\label{forward-model}
\begin{figure}
    \centering
    \includegraphics[width=86mm]{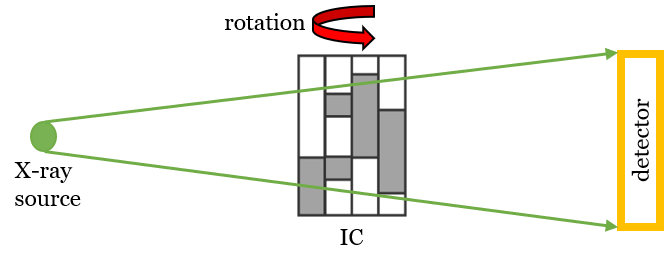}
    \caption{A conceptual diagram for our imaging system (IC as the object).}
    \label{fig:tomcat}
\end{figure}

An X-ray tomography imaging system usually consists of a sample placed between an X-ray source and a detector. 
The measurements are taken from a series of rotation angles of the sample, where a cone-beam geometry is assumed to produce the ray projection from the source point to the object, and then from the object to the center of each detector pixel. A conceptual diagram for the imaging system is in Fig.~\ref{fig:tomcat}.  Here the sample is a three-dimensional IC. In the absence of noise, the exact detection model is written as
\begin{equation} \label{exact}
\begin{split}
    g^{(0)} &= \int \! dE \, D(E) \, I^{(0)}(E) e^{ -\alpha(E)  Af}, 
\end{split}
\end{equation}
where $A$ is the system matrix, i.e., the distance each ray traverses from the source through the object to a detector pixel, with each column corresponding to one sample voxel and each row corresponding to one detector pixel over all sample rotation angles, $f$ is the vector of voxel compositions of the object (dielectric or copper for IC), $E$ is the photon energy, $\alpha(E)$ is the absorption coefficient at energy $E$ for copper, $I^{(0)}(E)$ is the source intensity, $D(E)$ is the detector efficiency, $g^{(0)}$ is the vector of expected number of photons measured for each of the detector pixels. The exponential is applied component-wise.
When the illumination is monochromatic, 
Eq.~\ref{exact} can be simplified to
\begin{equation} \label{forward}
\begin{split}
    g^{(0)} = N_0 e^{ -\alpha  Af},
\end{split}
\end{equation}
where $N_0$ is a vector containing the expected number of photons in each ray, i.e., the index of the element corresponds to a ray. $A$ is the system matrix, and $\alpha$ is the absorption coefficient as before. This can be written in the form of a linear equation

\begin{equation}
\begin{split}
   \ln g^{(0)} - \ln N_0  = -\alpha  Af.
\end{split}
\end{equation}
Here, the natural log is defined component-wise. In our forward simulations, we assume the measured photon counts are Poisson-distributed.

\section{Inverse algorithms for X-ray tomography}
\label{algo}

\subsection{Filtered Back-projection}
Filtered back-projection (FBP) is an analytical reconstruction algorithm that has been the default in X-ray tomography for years. It applies a filtering function to the measurements first and then produces the object reconstruction by back-projecting the measurements. The reconstruction quality from FBP method is sensitive to the sampling ratio and noise level in the Fourier space. When forward projections are in limited-angle and low photon conditions, FBP can produce unsatisfactory reconstructions containing noise and streak artifacts~\cite{frikel2013reconstructions}.

\subsection{Iterative algorithms with prior regularizer}
The Wiener-Tikhonov approach~\cite{kailath1981lectures, golub1999tikhonov} improves the analytic reconstruction by iteratively optimizing:
\begin{equation} \label{eq:optimazation}
    \hat{f} = \arg\min_f \big\{ \|-\alpha A f - \ln g + \ln N_0 \|^2 + \beta\, \Psi(f)\big\},
\end{equation}
where $\hat{f}$ is the inverse result, $\|\cdot \|$ is the $L^2$ norm, $\Psi(f)$ is the regularizer or Bayesian prior, and $\beta$ is a regularization parameter. Gaussian measurement noise is assumed in Eq.~\ref{eq:optimazation}. The optimization starts with an assumed object, simulates a set of measurements from the assumed object, compares the experimental and simulated measurements, and then updates the object based on the differences. The last step also includes the discrepancy in the prior term in the computation of the update. The process continues to iterate until a convergence criterion is achieved. The specific prior term $\Psi(f)$ is the key for artifact suppression and edge preservation~\cite{chen2008prior, heusser2014prior, schrapp2014data}, but such preoperative information is often difficult to acquire, or may even unavailable~\cite{zhang2016image}. Additionally, the value of the regularization parameter is a matter of importance yet is usually obtained by trial and error.

\subsection{Deep reconstruction network with learned prior}
Recently, deep-learning-based inversion has been applied to X-ray tomography. The approach, known as a deep reconstruction network, utilizes a prior distribution to generate high-quality reconstructions. The prior distribution can be learned using supervised training from a dataset consisting of pairs of ground truth and measurements thereof, it can also be the structure of a neural network (the inductive bias), e.g., a deep image prior. Here, we focus on methods with a learned prior distribution.

There are mainly two kinds of deep reconstruction networks with learned priors. The first kind is an End-to-End approach, where a direct mapping from the measurement to the object reconstruction is obtained by using measurement and ground truth object pairs as the training dataset~\cite{zhu2018image}. The network is trained without physics assisting, where it needs to implicitly learn the inverse physics and the object prior $\Psi(f)$ simultaneously. Reports have shown that End-to-End deep neural networks may conflate reconstructions whose difference lies either in or close to the null space of the system matrix, leading to instabilities of the reconstruction~\cite{antun2020instabilities, gottschling2020troublesome}. 
This suggests the need to instead use a second kind of network with some level of physics assisting, in which the reconstruction is a two-step process. The first step is to use FBP algorithm to produce a noisy reconstruction based on the known imaging geometry from the measurements, and the second step is to use a deep network to remove the noise and artifacts within the FBP reconstruction~\cite{chen2017low, jin2017deep, yang2018low, liu2020tomogan}.  This way, the network only learns the prior $\Psi(f)$ from the FBP reconstruction and ground truth object pairs without considering the inverse physics from measurement to object, achieving good computational efficiency during inference. Some studies replace the FBP algorithm in the first step using another deep reconstruction network~\cite{he2020radon, wu2020stabilizing} to overcome the issue of the instabilities.

\section{Physics-assisted Generative Adversarial Network (PGAN)}
\label{PGAN}

Our proposed PGAN improves upon the second kind of deep reconstruction network using the two-step reconstruction process. Rather than an FBP reconstruction, we utilize a maximum-likelihood estimation resulting from an iterative algorithm with known physics to map the measurement to an approximant object. The physics that we know \textit{a priori}, i.e., the forward imaging geometry of the X-ray tomography and the Poisson noise in low-photon measurements, are incorporated in the first step of the reconstruction. Then, a generative model uses the learned prior $\Psi(f)$ to further improve the maximum-likelihood estimate. Therefore, the PGAN inversion framework incorporates the known physics from an iterative algorithm 
and the learned prior from a deep generative model, drastically improving the reconstruction quality. The details of our algorithm are presented below.

\subsection{Maximum-likelihood estimate}

Maximum-likelihood estimates are tomographic reconstructions obtained by maximizing an objective function based on a likelihood which incorporates the projective geometry and Poisson statistics.  These reconstructions serve both as the input to our generative models as well as being a standard algorithm to which we compare our machine learning results.
The method produces the maximum log-likelihood reconstruction $\Tilde{f}$ with a given set of tomographic measurements $g$ (in the number of photon counts for each detector pixel) assuming that the measurement noise is consistent with a Poisson process. The objective is to find an optimal $\Tilde{f}$ given the measurements $g$ by solving
\begin{eqnarray}
\label{eq:optimization}
\Tilde{f}(g) &=& \arg\max_{f^{(0)}} \left[ L_{\rm MLE}(g | f^{(0)}) + \Psi(f^{(0)})\right] \ \ \ \ \ \textrm{and}\\
     L_{\rm MLE}(g | f^{(0)}) &=&
 -\sum_{i} \left[ \ln g_i! - g_i \ln g_i^{(0)} + g_i^{(0)} \right].
\end{eqnarray}
Here $L_{\rm MLE}$ is the log-likelihood under the assumption of Poisson statistics \cite{sauer_bouman1993}, $\Psi$ is a regularization function or log of the Bayesian prior, $g^{(0)}$ is the simulated measurement from a proposed object $f^{(0)}$ based on Eq.~\ref{forward} without considering noise statistics, $\sum_i$ sums over all individual measurements where $i$ indexes the measurements at different angle-detector pixel pairs, $\ln g_i!$ takes the log of a particular measurement, and $\Tilde{f}$ is the optimal reconstruction based on maximizing the log likelihood~\cite{Levine2019,Levine2021,Bouman1993,sauer_bouman1993}.
More information is given in Appendix~\ref{app:MLEinfo}.
\subsection{Deep generative models}
\label{deep networks}
\begin{figure}
\centering
\begin{subfigure}[b]{0.55\textwidth}
  \includegraphics[width=1\linewidth]{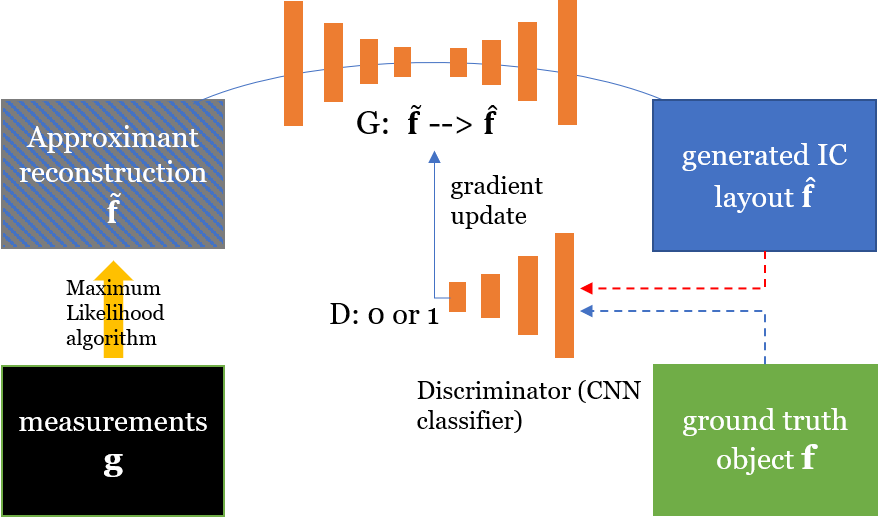}
  \caption{}
  \label{fig:GANa} 
\end{subfigure}

\begin{subfigure}[b]{0.55\textwidth}
  \includegraphics[width=1\linewidth]{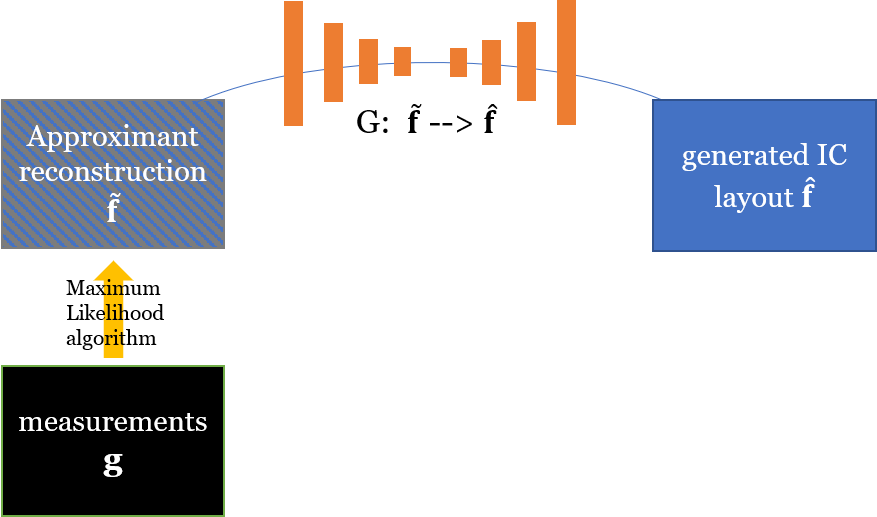}
  \caption{}
  \label{fig:GANb}
\end{subfigure}

\caption[generative framework]{(a) Supervised training of the generative model using pairs $\left(\tilde{f},g\right)$. (b) Testing on pairs never used during training.}
\end{figure}

Our deep generative model is based on a supervised machine learning technique known as conditional generative adversarial network (cGAN)~\cite{mirza2014conditional}. The generative model learns a prior distribution of the object, and then improves the 3D reconstruction from maximum-likelihood estimation. Here, the maximum-likelihood estimation is the conditional information for the deep generative model. When the available projection angles and photons per ray are limited for X-ray tomography, the reconstructions from maximum-likelihood estimation contain artifacts due to the missing cone problem.  The quality of these reconstructions will eventually drop below an acceptable threshold when the angular range or photon flux of the tomographic measurements decreases. Based on our numerical experiments, the deep generative model improves the noisy maximum-likelihood reconstructions resulting in an output object structure which better replicates the true object. Because the generative model is conditioned on the reconstructions knowing the imaging geometry of the system and the Poisson statistics in the measurements, the learned prior can be more effective than those methods using conditional information without or with less physical priors.


In GAN's original form, the objective is to map a random vector $z$ to the targeted distribution given a set of samples from the true distribution $f$:
\begin{equation}\label{eq:GAN}
\arg \min_{G} \max_{D} \Big(\mathbb{E}_{f} \big[ \log D(f) \big] + \mathbb{E}_{z} \big[ \log \big(1 - D(G(z))\big) \big] \Big),
\end{equation}
where $G$ is the generator and $D$ is the discriminator. 
The optimization process is a competition between $G$ and $D$, where the generator tries to create examples as realistic as possible to deceive the discriminator while the discriminator tries to distinguish generated examples from the given true examples~\cite{goodfellow2014generative}. Therefore, the generator tries to minimize the objective function and the discriminator tries to maximize it. The cGAN method has achieved impressive results not only in computer vision~\cite{yangjie2018review}, but also in physics-related applications, including computer-generated holography~\cite{HosseinEybposh:20}, medical imaging~\cite{YI2019101552}, and the solution of differential equations~\cite{mgansalik2020learning}. Conditional GAN is an extension to the original GAN model: it modifies the original GAN by conditioning both the generator and discriminator on some extra information about the distribution we try to synthesize. Whereas the original GAN has a problematic instability in training~\cite{mescheder2018training}, cGAN gives us control over modes of the distribution to be generated. This is why we choose cGAN as the generative model. Our intuition is that when more physical priors are included in the conditional information, a better controlled generative process can yield higher qualities of reconstructions.

 In our case, the conditional information for the deep generative model is the noisy maximum-likelihood approximant $\Tilde{f}$. This is the sole input to the generative model, and we do not include the a random vector. The training distributions are the pairs of approximant $\Tilde{f}$ and ground truth $f$. Note that $f^{(0)}$ introduced previously represents a proposed reconstruction.
The objective involving the discriminator of our generative model is formulated by modifying the original cGAN's objective~\cite{mirza2014conditional}, and is:
\begin{equation}\label{eq:cGAN}
\arg \min_{G} \max_{D} \mathbb{E}_{\left(f,\tilde{f}\right)} \big[ \log D(f) + \log \big(1 - D(G(\Tilde{f}))\big) \big].
\end{equation}
 The complete objective function of our generative model, including the supervised term, can be found in the Appendix~\ref{app:train}. Note that the expectation is approximated in our simulation by using the arithmetic mean over a batch of training samples. Through training, the generative model learns a prior via the competition between the generator and the discriminator. When the competition reaches the Nash equilibrium, the training process is complete. 
 We denote the output of the trained generative model as $\hat{f}$. 
 
 

The generator in our model is a 3D autoencoder that first learns to convert an object representation to a latent space representation using an encoder module, and then decodes the representation back to 3D object. The discriminator is a 3D convolutional classifier that tries to find whether the output from the generator is realistic or not (reporting a floating-point number in the range of [0, 1]). The generator and the discriminator have convolutional kernels that are spectrally normalized to stabilize the training process~\cite{miyato2018spectral}. The discriminator is updated with the generator only during training and is not needed during testing. Fig.~\ref{fig:GANa} shows the training process for our deep generative model, and Fig.~\ref{fig:GANb} shows the process of testing and inference. In total, four variants of the deep generative model are investigated, in particular: a deep generative model (baseline model), a generative model with axial attention, a generative model with a scattering representation, and a generative axial model with a scattering representation. 
The variants alter the design of the encoding module while all sharing the same decoding modules and discriminator architecture. Detailed architectures can be found in the Appendix, and
online~\cite{Guo_Physics-assisted_Generative_Adversarial_2021}.

The baseline generative model uses a series of cascaded 3D convolutional layers in alternation with pooling layers in the encoder module to extract features from the input reconstruction. Increasing the number of convolutional layers in the encoder can enable the module to learn more complicated features from the input~\cite{ba2013deep}, capturing high-level spatial dependencies of training objects.

Our second variant of the generative model is based on axial attention that harvests the contextual information in the input approximant. To build such an encoder module, we replace some of the 3D convolution layers in the encoder of the generator with full axial attention to extract or detect global features in the input reconstructions. 
The core idea of this technique is to factorize the 3D self-attention
into three 1D attention modules along the height, width, and depth axis sequentially, which can reduce the computational complexity of 3D self-attention to $\mathcal{O}(hwzm)$, where $h$, $w$, $z$, are the height, width, and depth of the input features, respectively, and $m$ is the local constraint constant~\cite{wang2020axial}. Therefore, axial attention is more efficient than the standard self-attention, enabling long-range and global feature learning to overcome the limitation of locality in the convolutional kernels. In our implementation, the local constraint is set to be the same size as the given axis, and each axial attention has eight attention heads. 


The third and fourth variants of generative models include the wavelet scattering transform~\cite{anden2014deep, mallat2012group} of the reconstruction as an additional input to the encoder module of the model. A scattering representation can be produced without training, and such a representation is capable of including features at multiple scales. When combined with the renormalization technique, the generative model can be further conditioned on the scattering representation in generating realistic objects. In particular, after being fed to a trainable transformation of a fully connected layer, the wavelet representation re-scales and re-shifts the normalized feature values from convolution or axial attention. 
This technique may provide supplementary features of the input approximant into the reconstruction process of the neural network~\cite{delbrouck2017modulating}, which may help the network in learning the mapping from noisy reconstruction to noiseless reconstruction. 

\section{Evaluation methods}
\label{eval}
\subsection{CircuitFaker for tomographic objects}
\label{CircuitFaker}
\begin{figure}
    \centering
    \includegraphics[width=86mm]{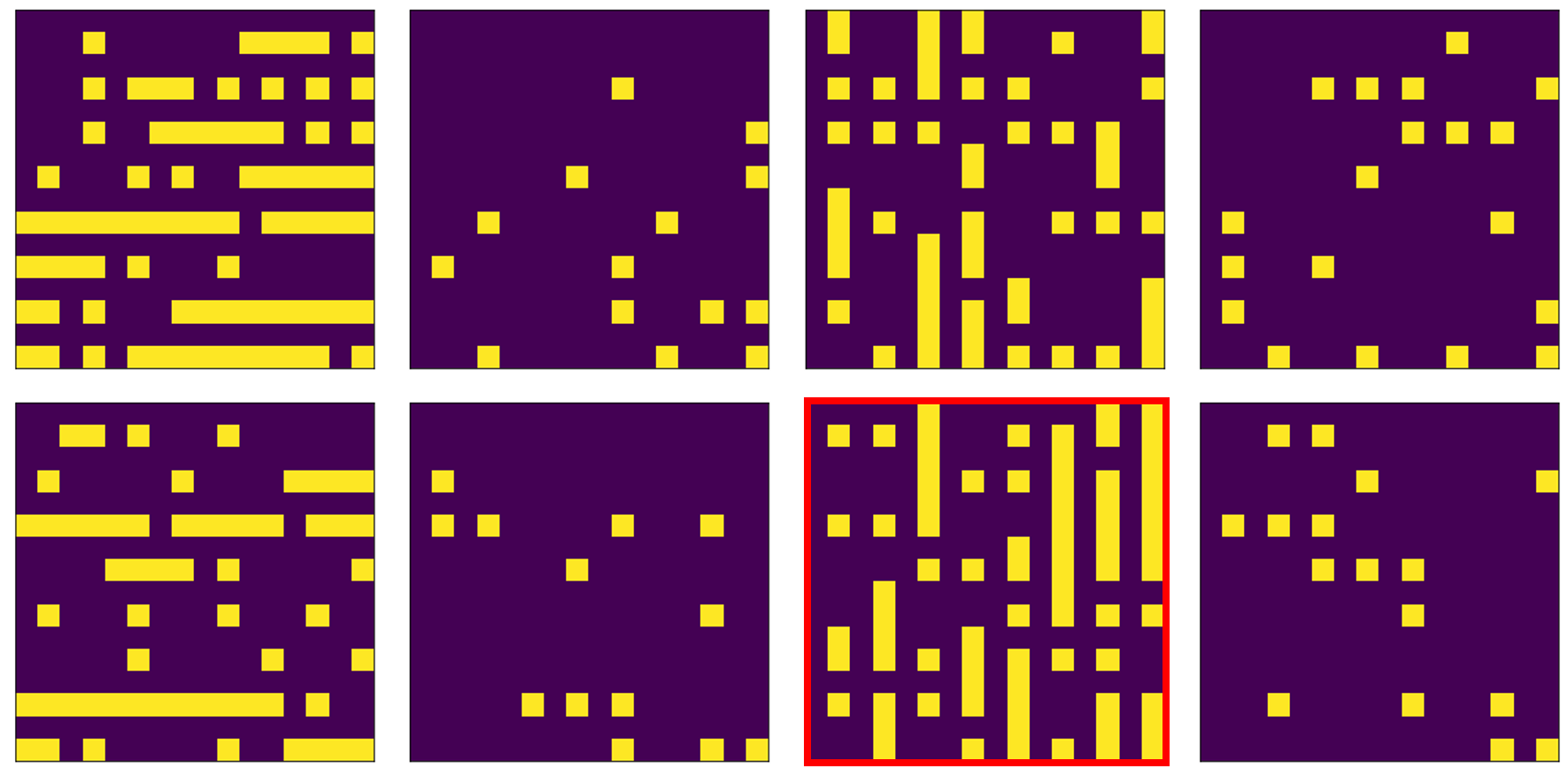}
    \caption{Selected $16\times16\times8$ circuit from CircuitFaker. Each image is a slice of 2D layer in the $z$ dimension. The value of $z$ increases as a raster scan of the 8 slices shown.  Yellow indicates copper and purple indicates silicon. Here, $x$ layers are the first (upper left) and fifth layers (lower left) in $z$, $y$ layers are the third and seventh layers in $z$. Others are via layers. The layer highlighted in red is the ground truth circuit layer in the later comparisons.} 
    \label{fig:expcircuit}
\end{figure}

CircuitFaker is an algorithm that generates a random set of voxels with binary values resembling an integrated circuit interconnect. The synthetic circuits from CircuitFaker form the class of artificial objects for tomographic reconstruction, and the implicit correlations in their spatial features are the priors to be assumed or to be learned for the inverse algorithms. A particular draw of CircuitFaker assigns a bit in each of $N = N_x N_y N_z$ locations.
These locations are indexed as $i_{\ell} = 1, ..., N_{\ell}$, with $\ell$ = $1$, $2$, $3$ for $x$, $y$, and $z$. All bits are initialized to $0$. In the first round, there are wire seed points for all locations ($i_1$, $i_2$, $i_3$) with $i_1$, ..., $i_3$ odd. Each seed point is set by a Bernoulli draw with probability $p_w$ of getting a $1$.
There are three kinds of layers, $x$, $y$, and via layers. The $x$ wiring layers have index $i_3 = 1$ mod~4. The $y$ wiring layers have $i_3 = 3$ mod~4. The via layers are the others, i.e., $i_3$ even.
In the second round, a point on an $x$ wiring layer to the immediate right of a point with value~1 is set to~1 with probability $p_x$.  A point on a $y$ wiring layer immediately above in plan view a point with a value~1 is set to~1 with probability $p_y$.  Similarly, a point on a via layer immediately above a point with a value 1 is set to~1 with probability $p_z$.
In this paper, we chose these parameters: $N_x=N_y=16$, $N_z=8$, $p_w=0.75$, $p_x=p_y=0.8$, and $p_z=0.5$. Fig.~\ref{fig:expcircuit} shows one of the generated circuits with size $16\times16\times8$. 

\subsection{Imaging geometry for X-ray tomography}
\label{X-ray}
The imaging geometry is chosen to support an experimental project to perform integrated circuit tomography with a laboratory-scale instrument~\cite{Szypryt2021}. Each voxel in the circuit is of size 0.15~$\mu$m $\times$ 0.15~$\mu$m $\times$ 0.30~$\mu$m to emulate a real-world circuit. Therefore, the total volume of the circuit is 2.4~$\mu$m $\times$ 2.4~$\mu$m $\times$ 2.4~$\mu$m, and the voxel size is total of $16\times16\times8$. The detector is assumed to be in the $x$-$z$ plane at a tilt angle of $\varphi=0^\circ$.  The rotation axis is $z$.  The detector is 13.44~mm $\times$ 13.44~mm with 32 $\times$ 32 pixels of size 420~$\mu$m $\times$ 420~$\mu$m.  The system operates with a geometric magnification of 5000, with a source-sample distance of 10~$\mu$m.
There are eight tilt angles from $-30^\circ$ to $+22.5^\circ$ with an increment of $7.5^\circ$.  There is a single source point with a cone-beam geometry.  A single ray is taken from the source point to the center of each detector pixel. Minor corrections for variations in the source-to-detector pixel distance, the obliquity, and the source's Heel effect are neglected. In this work, we do not use scatter corrections, and we are restricted to a single material, namely copper, at its bulk density of 8.960~g/cm$^3$. Therefore, the reconstruction at each voxel ends up being a binary variable. The spectrum consists of two equally weighted lines at 9362~eV and 9442~eV, the Pt L$_\alpha$ fluorescence lines.  The attenuation per voxel is about {2\%} if copper is present.  The exact value depends on the details of how a ray intersects a voxel.

\subsection{Bit-error-rate formulation}
\label{BER}
The bit error rate (BER) is introduced as an evaluation metric to assess the performance of the reconstruction quality. It provides a measure of the frequency of misclassification for binary values in the voxels in a given circuit. That is, BER is the probability of classifying a specific voxel in a circuit to be 1 while the ground truth value for the corresponding voxel is 0 and \textit{vice versa}. The procedure for computing bit error rate in this paper is slightly modified from the standard used in communication theory, and is as follows:
\begin{enumerate}
    \item Compute posterior distributions $p(f_i = 0 \,|\, \Tilde{f})$
by multiplying the probability density functions (PDFs) $p(\Tilde{f}\,|\,f_i=0)$ and $p(\Tilde{f}\,|\,f_i=1)$ and their corresponding prior distributions ($p_0$ and $p_1$ for $f$). Here, $f_i$ represents an individual voxel in the circuit.
\item Apply a threshold based on a likelihood function to classify 0 and 1, where the intersection of the distributions of 0 and 1 determines the threshold. In our implementation, the prior likelihood functions are normal distributions. 
\item Compute the error rates for 0 and 1 ($\eta_0$ and $\eta_1$, respectively) by summing over the misclassified region in the probability density functions. 
\item Derive the expected bit error rate:
$\eta_{\rm avg}=\eta_0 p_0+\eta_1 p_1$.
\end{enumerate}

\section{Numerical results}
\label{numerical}
\subsection{Reconstructions for IC}
To demonstrate the ability of the learned prior from the generative model, we investigate its performance in solving ill-conditioned tomography problems. First, we show the advantage of the physics-assisted approach by comparing our PGAN to methods with no physics-assisting and less physics-assisting. The imaging conditions are constrained to limited angle and low-photon cases where the effect of ill-conditioning becomes severe. The approach without physics assisting is End-to-End training, where the generative model directly transforms the X-ray projections to the imaging object without knowing the imaging geometry and the Poisson statistics in the measurements. The conditional information for the generative model is the physical measurements. The strategy with less physics assisting is to use Filtered Back Projection as the conditional information for the generative model to only include the known imaging geometry without knowing the Poisson statistics in the measurements. In contrast, PGAN include the known imaging geometry as well as the Poisson statistics in the measurements by using the maximum-likelihood estimates (MLE) as the conditional information. The generative model (using the baseline model) and its training parameters are identical across the three approaches for a fair comparison, except for End-to-End method where a slight adaptation of dimension in the network architecture is necessary and applied. Detailed information about the adaptation can be found in Appendix~\ref{app:networkArch}. 

\begin{figure}
    \centering
    \includegraphics[width=0.8\textwidth]{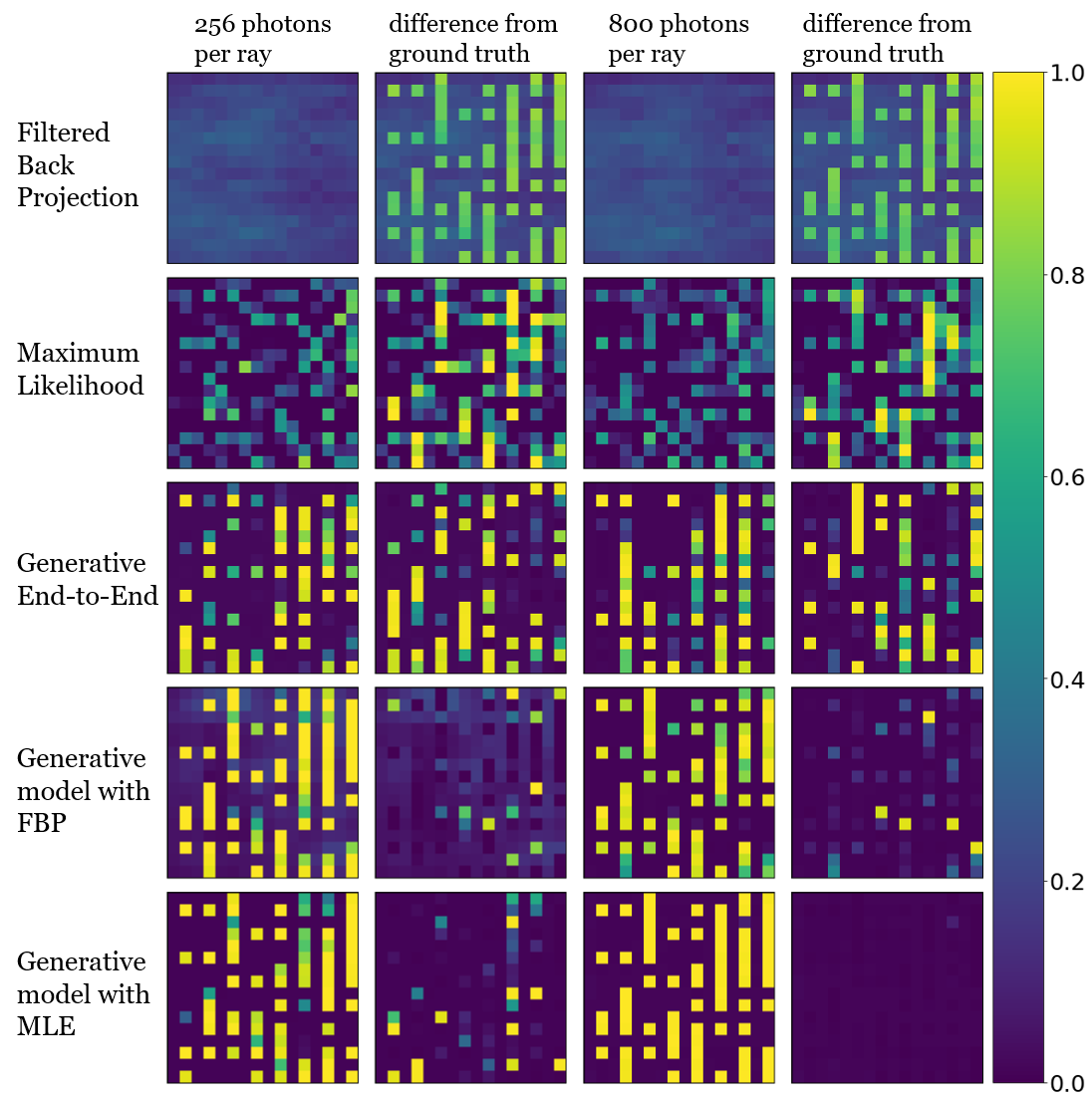}
    \caption{Selected examples of IC reconstructions with an angular range of $-30^{\circ}$ to $22.5^\circ$. The color scale runs from 0 to~1. Each row represents different reconstruction methods, and each column in odd number represents the same location at the given IC distribution with a different photon number per ray, and each column in even number shows the absolute difference between the reconstruction in the previous column to the ground truth circuit.}
    \label{fig:ic-compare-complete}
\end{figure}

Fig.~\ref{fig:ic-compare-complete} shows selected examples of IC reconstructions in limited-angle and low-photon conditions. Each of the simulations for the generative model is done with 1800 training sets and 200 test sets, with each set having $16\times16\times8$ voxels. Two photon conditions (256 and 800 photons per ray) are showed. Each row represents different reconstruction methods, and each column in odd number represents the same location at the given IC distribution with a different photon number per ray, and each column in even number shows the absolute difference between the reconstruction in the previous column to the ground truth circuit. As the total photon increases, the IC reconstruction qualities for all the methods using generative models increase, though visually, reconstruction qualities of MLE and FBP methods hardly improve. More importantly, for methods with generative models, using MLE as the conditional input achieves the best reconstruction quality. Generative model with FBP is slightly under while End-to-End method has the most errors. This demonstrates the advantage of our physics-assisted approach. The more physical priors included as the conditional information for the generative models, the less difference there will be between the IC reconstruction and ground truth. 

\begin{figure}
    \centering
    \includegraphics[width=86mm]{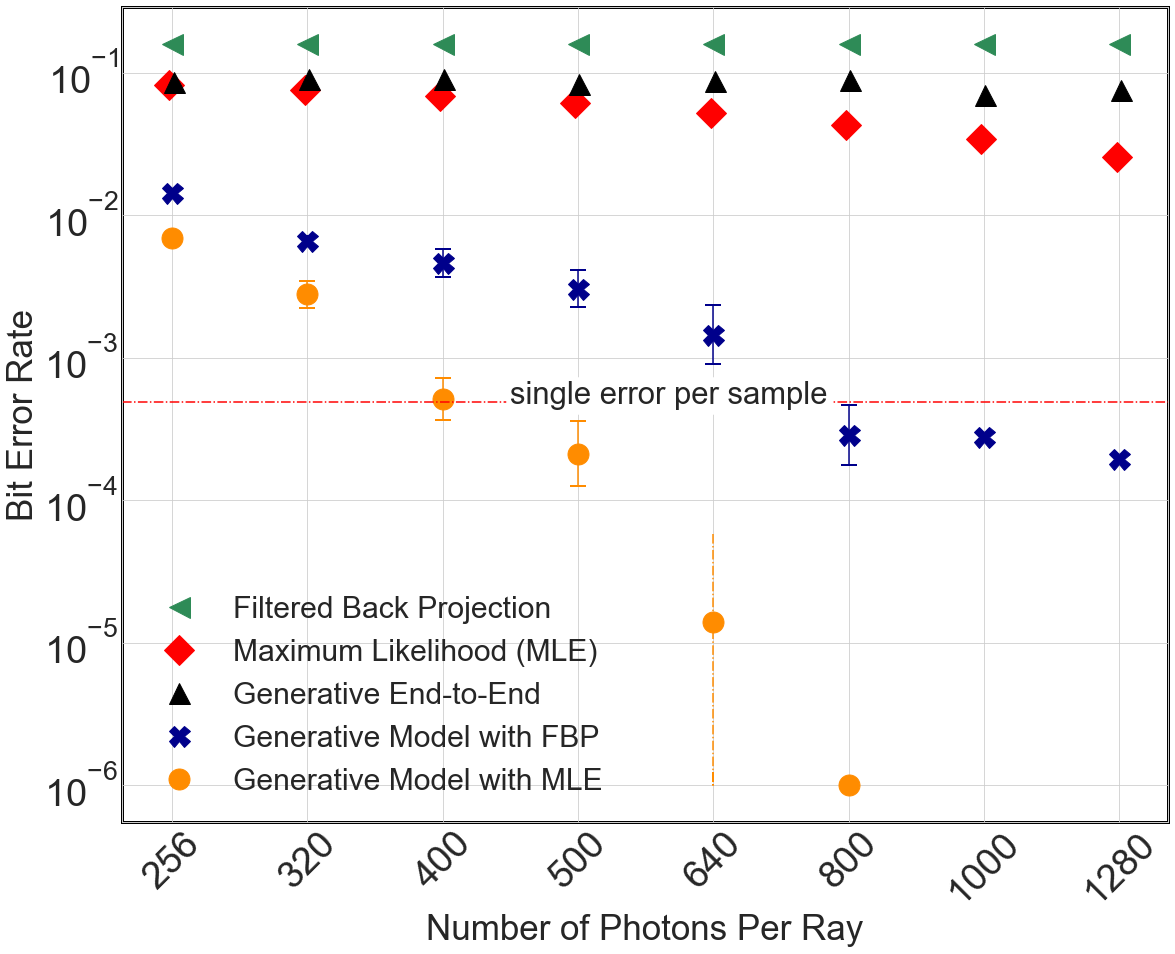}
    \caption{Different reconstruction approaches with an angular range of $-30^{\circ}$ to $22.5^\circ$ for IC objects. Single-error-per-sample in the red dashed line means that the bit error rate is equal to $\frac{1}{16\times16\times8}$.}
    \label{fig:MLvs2}
\end{figure}

The quantitative comparison of different approaches is shown in Fig.~\ref{fig:MLvs2}. The $x$~axis is the number of photons per ray, the $y$~axis is the bit error rate of the reconstructed 3D IC test dataset. For the transition cases between 320 and 800 photons per ray, we repeat the simulation with a total of five independent synthetic sets of IC circuits and report the means and standard errors in the plot. The results agreed with our observation in Fig.~\ref{fig:ic-compare-complete}. Generative model with MLE achieves single-error-per-sample only at 400 photons.
Generative model with FBP is the second best requiring around 800 photons for single-error-per-sample. Generative model with End-to-End training, however, only shows limited improvement with increasing photons.

To evaluate the design choice of our generative model, we extended our study in a larger range of photon number per ray. Now, variants of the generative models share the same conditional information (maximum-likelihood reconstructions with the known physical priors) and training parameters, with differences in network architectures.

\begin{figure}
    \centering
    \includegraphics[width=0.8\textwidth]{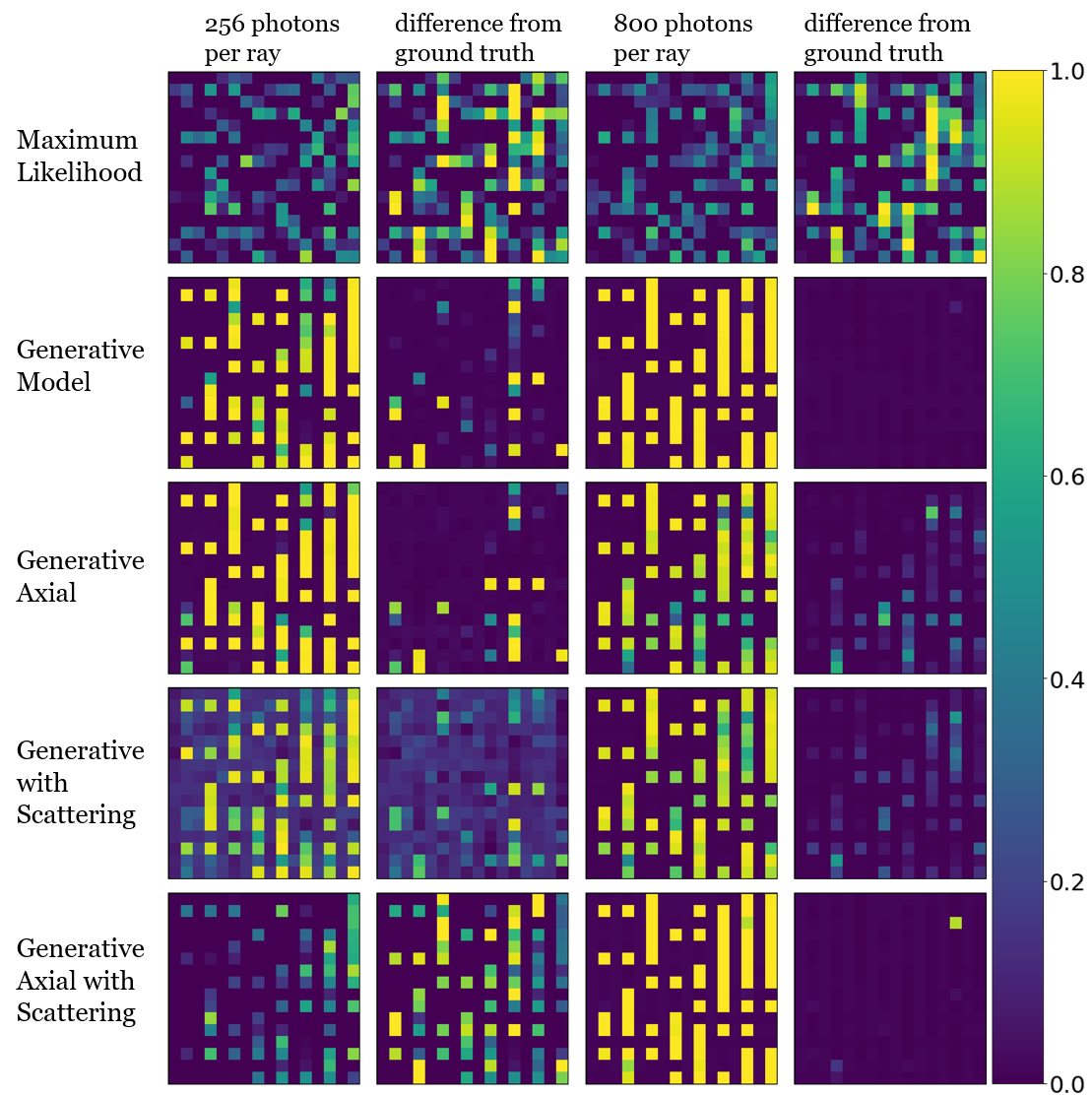}
    \caption{Selected examples of IC reconstructions with an angular range of $-30^{\circ}$ to $22.5^\circ$. The color scale runs from 0 to~1. Each row represents different reconstruction methods, and each column in odd number represents the same location at the given IC distribution with a different photon number per ray, and each column in even number shows the absolute difference between the reconstruction in the previous column to the ground truth circuit.}
    \label{fig:ic-compare}
\end{figure}

Fig.~\ref{fig:ic-compare} shows selected examples of IC reconstructions in the limited-angle and low-photon conditions as before. Each row represents different reconstruction methods, and each column in odd number represents the same location at the given IC distribution with a different photon number per ray, and each column in even number shows the absolute difference between the reconstruction in the previous column to the ground truth circuit. Considerable improvement is visible when comparing the maximum-likelihood reconstructions and generative reconstructions.

\begin{figure}
    \centering
    \includegraphics[width=86mm]{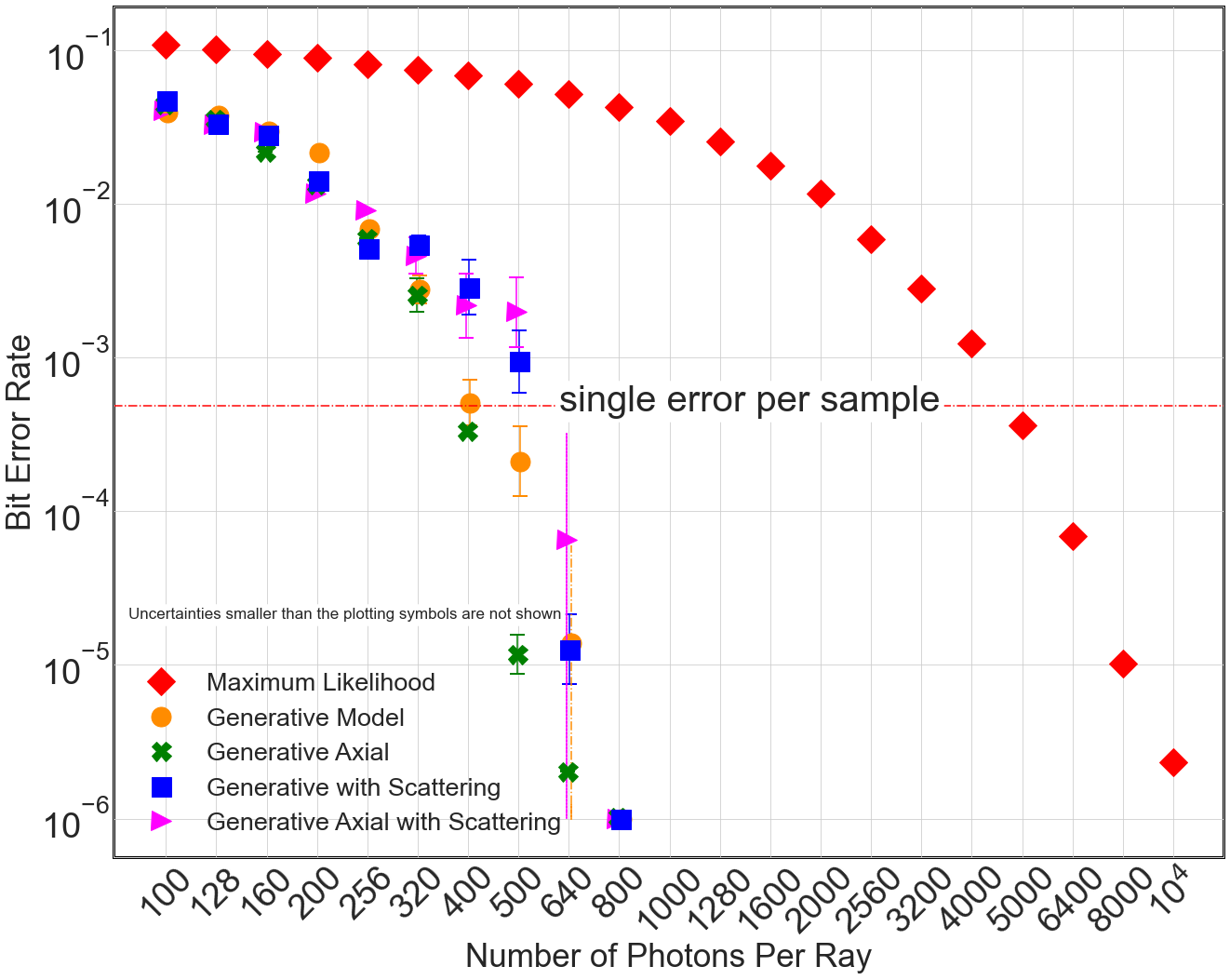}
    \caption{Maximum-likelihood vs. generative model reconstructions with an angular range of $-30^{\circ}$ to $22.5^\circ$ for IC objects.}
    \label{fig:MLvsGANb}
\end{figure}

The quantitative comparison of limited angle and low-photon tomography is shown in Fig.~\ref{fig:MLvsGANb}. The $x$~axis is the number of photons per ray in the tomographic projection which ranges from 100 to $10^4$. The $y$~axis is the averaged bit error rate of the reconstructed 3D IC test dataset.
For generative models, the transition from above a single-error-per-sample to below happens between 320 to~640 photons per ray. 
For the transition cases between 320 and 640 photons per ray, we repeat the simulation with a total of five independent synthetic sets of IC circuits and report the means and standard errors in the plot.  With 640 photons per ray, the bit error rates from generative model reconstructions drop at least two orders of magnitude relative to the maximum-likelihood reconstructions. In particular, the generative model with axial attention performs the best in terms of its lower mean and standard error,  reaching a single-error-per-sample at 400 photons per ray. We may attribute this to the application of axial attention in capturing long-range interactions within the input. Generative models with wavelet scattering representation show no advantage in performance. This may be due to the small size of the input, where the additional information from the wavelet representation may have been learned from convolutional kernels and axial-attention modules. The maximum-likelihood reconstructions reach a single-error-per-sample when the number of photons per ray is around 5000. Therefore, in simulation, the generative models can reduce the photon number to reach a single-error-per-sample by one order of magnitude.

\begin{figure}
    \centering
    \includegraphics[width=86mm]{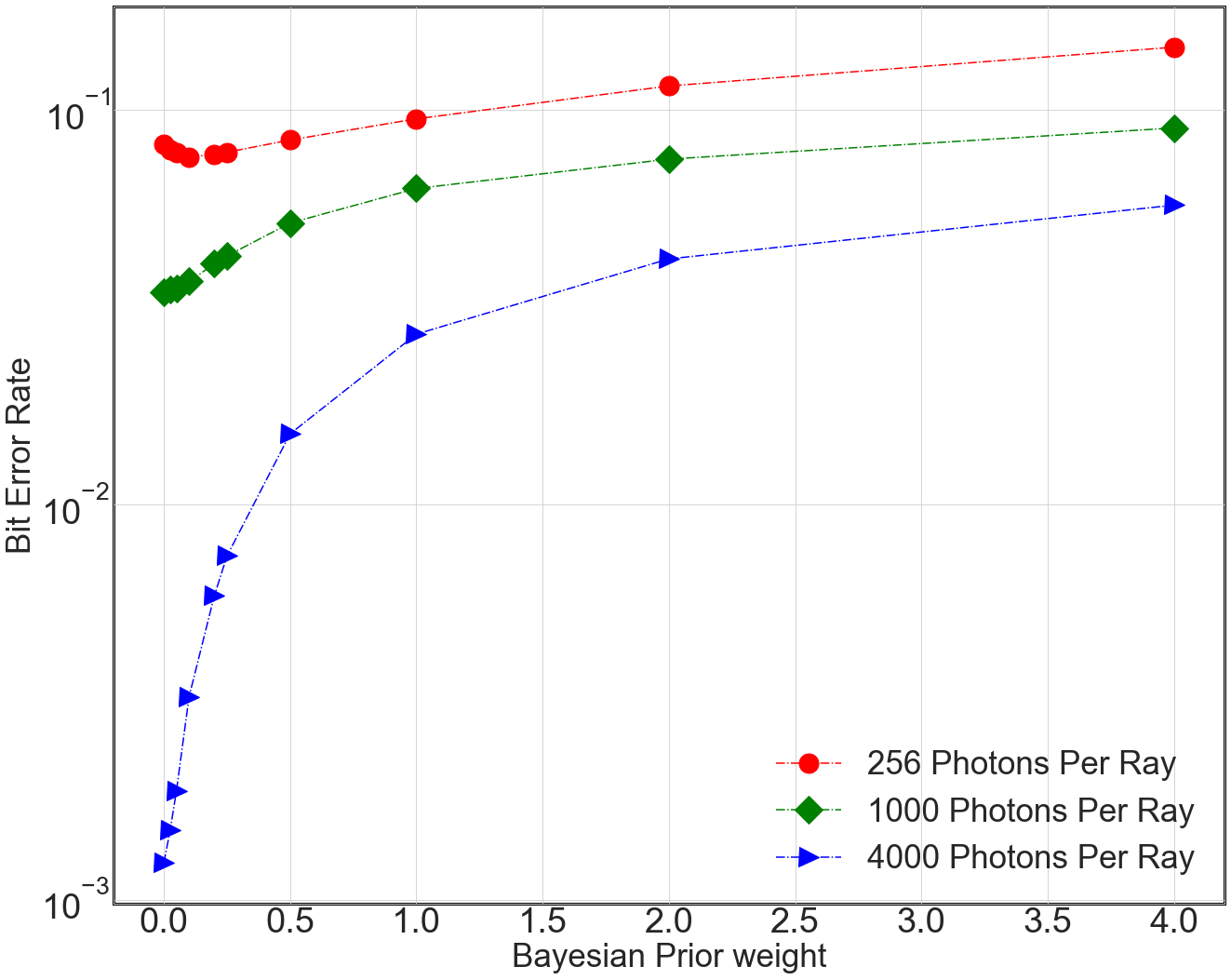}
    \caption{Maximum-likelihood reconstructions including the Bouman-Sauer prior with an angular range of $-30^{\circ}$ to $22.5^\circ$ for IC objects.}
    \label{Bayprior}
\end{figure}

To demonstrate the challenges of using Bayesian prior in the iterative algorithm, and the advantages of the learned prior, we include a Bouman-Sauer prior in the maximum-likelihood estimate as an example. The selected Bouman-Sauer prior imposes the smoothness of the reconstruction. It is very similar to the Total Variation (TV) prior, particularly since we choose the weights of the neighbors to mimic the absolute value of the gradient~\cite{Levine2021}. Fig.~\ref{Bayprior} shows the reconstruction quality versus the regularization weight for three imaging conditions with limited angles. The measurements are the 200 test datasets with 3D ICs as the sample, identical to dataset that generates Fig.~\ref{fig:MLvsGANb}. Note that when the weight is~0, the reconstruction is the effectively same as maximum-likelihood estimate without the prior (our baseline in the main text). The Bouman-Sauer prior provides limited improvement for 256 photons per ray case. Quality degradation is obvious for the cases of 1000 and 4000 photons per ray. The results indicate that one particular Bayesian prior from the literature is not helpful for our problem.

\subsection{Reconstructions for independent coin toss object}

\begin{figure}%
    \centering
    \subfloat[\centering $p=0.5$ Bernoulli trial (fair coin)]{{\includegraphics[width=0.46\textwidth]{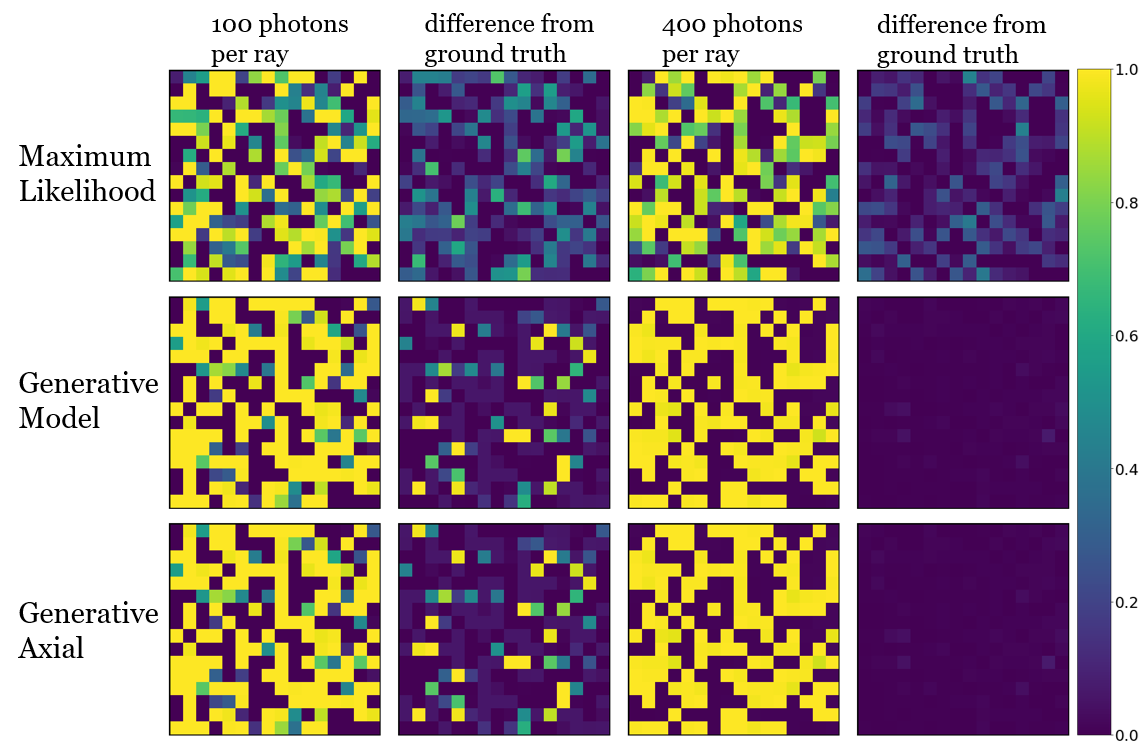} }}%
    \qquad
    \subfloat[\centering $p=0.18521$ (unfair coin matching circuit fill factor)]{{\includegraphics[width=0.46\textwidth]{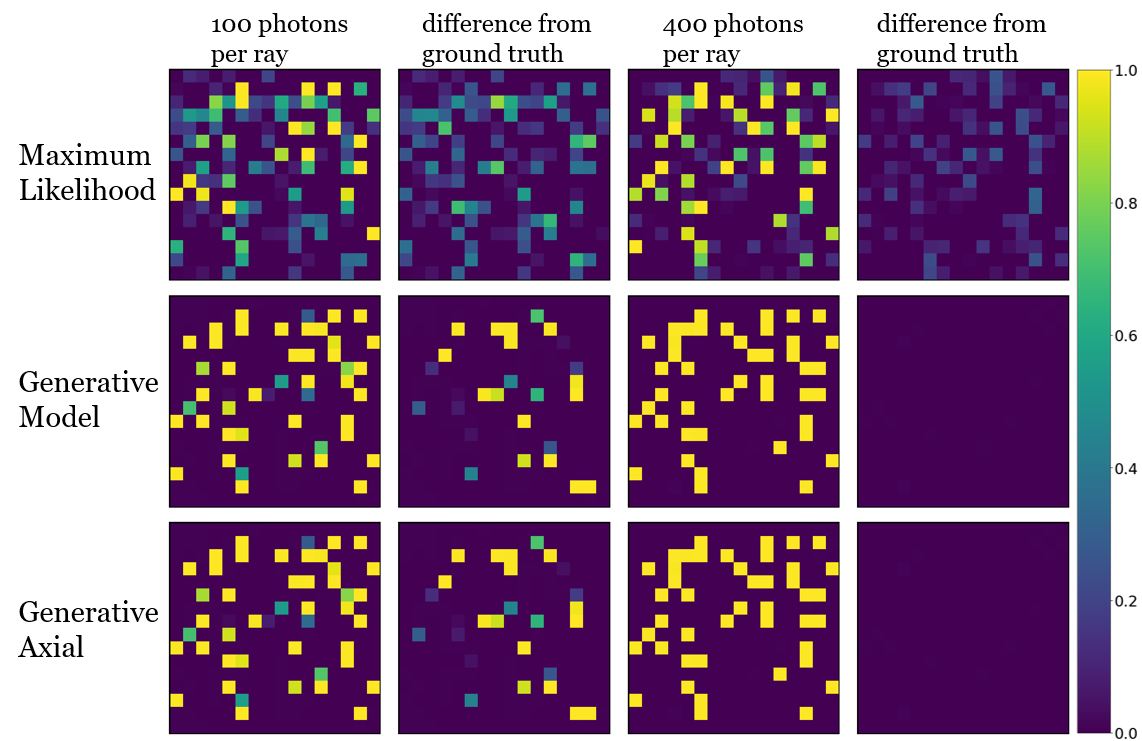} }}%
    \caption{Selected examples of independent coin toss an angular range of $-30^{\circ}$ to $22.5^\circ$. The color scale runs from 0 to~1. Each row represents different reconstruction methods, and each column in odd number represents the same location at the given IC distribution with a different photon number per ray, and each column in even number shows the absolute difference between the reconstruction in the previous column to the ground truth circuit. Reconstructions for Generative Axial at 400 photons per ray resemble the ground truth circuit layer.}%
    \label{fig:expcointoss}%
\end{figure}

To confirm that the improvement from the generative approach may indeed be attributed fairly to the learned prior, we further demonstrate the quantitative comparison of limited angle and low-photon tomography on 3D objects that are not spatially correlated. These 3D objects that are generated with an independent coin toss at every voxel. The probability of being 1 (copper) is 0.5 for fair coin toss, or unfair coin toss that matches the fill fraction for CircuitFaker generated circuits (which is $p=0.18521$, showing 1 standard deviation of statistical uncertainty). The learned prior in these cases is the probability $p$ for each voxel. Since the voxels are not spatially correlated, the learned prior from the generative model is expected to be less effective in solving the inverse problem. 

Fig.~\ref{fig:expcointoss} shows the selected examples of independent 3D object reconstruction by limited angle and low-photon tomography. The imaging geometry is the same as before, where the angular range is fixed at $-30^\circ$ to $22.5^{\circ}$ with $7.5^{\circ}$ steps. 
The improvement from the deep generative model is less pronounced than having a circuit object. 
\begin{figure}%
    \centering
    \subfloat[\centering $p=0.5$ Bernoulli trial (fair coin)]{{\includegraphics[width=0.46\textwidth]{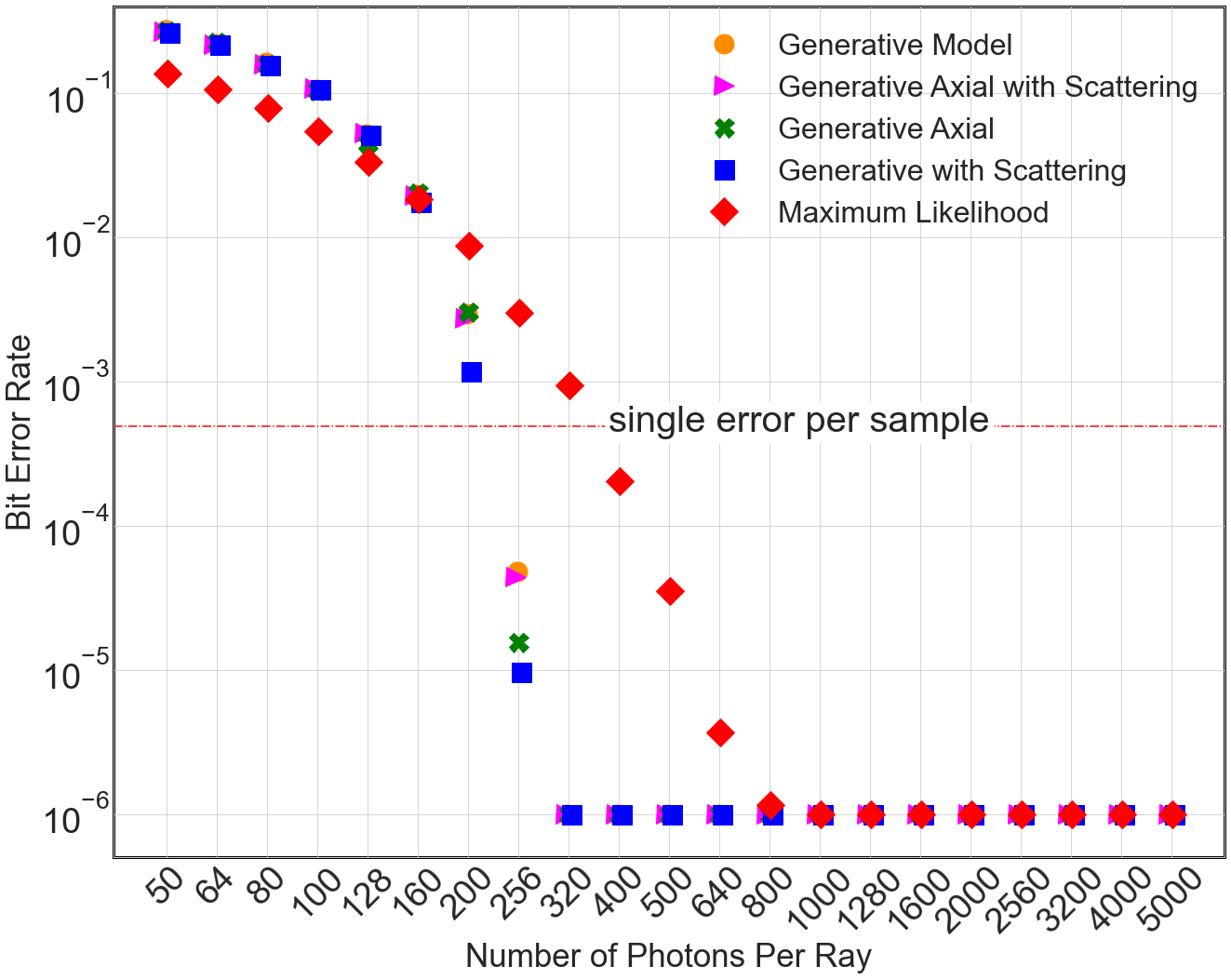} }}%
    \qquad
    \subfloat[\centering $p=0.18521$ (unfair coin matching circuit fill factor)]{{\includegraphics[width=0.46\textwidth]{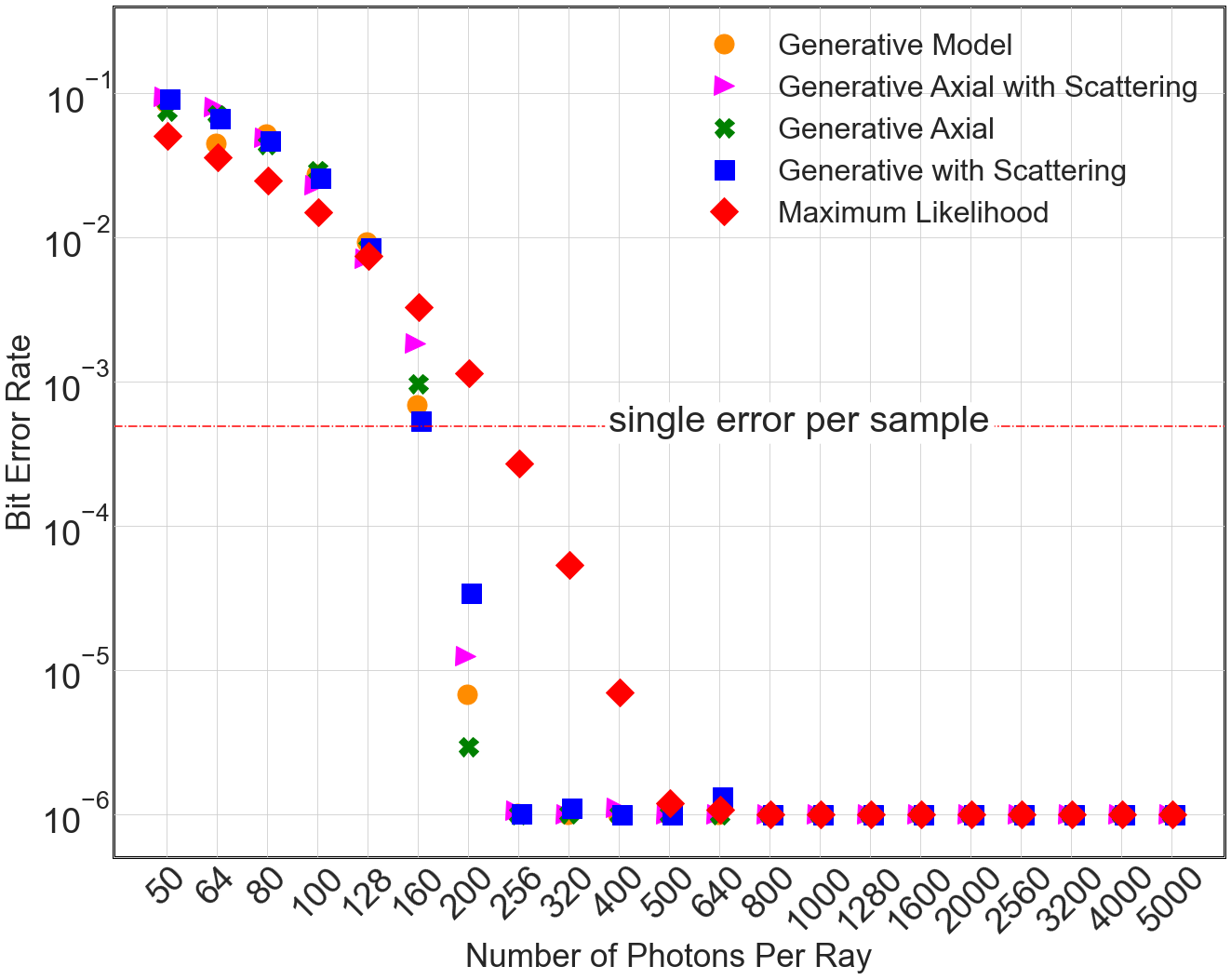} }}%
    \caption{Results for independent coin toss at every voxel with an angular range of $-30^{\circ}$ to $22.5^\circ$.}%
    \label{fig:cointoss}%
\end{figure}
The assumed prior (namely, the known physics and that each voxel has a value in $[0, 1]$) in our maximum-likelihood approach is now more proper to the reconstruction object with independent voxel. Therefore, the maximum-likelihood estimate improves. The learned prior from the deep generative model behaves similarly to a better classification cut-off for each voxel: generative models may predict each voxel value centered around 0 and~1 since all the 3D objects for training are binary. The maximum-likelihood approach does not assume objects are binary and it may produce reconstructions with more significant variations to the mean at the same imaging condition.

Fig.~\ref{fig:cointoss} shows the quantitative comparison for independent 3D object reconstructions. The $x$ axis is the number of photons per ray in the tomographic projection ranging from 50 to $5000$, $y$ axis is the averaged bit error rate of the reconstructed 3D coin toss test dataset. Fig.~\ref{fig:cointoss}(a) is the case with fair coin toss. Compared to the case with spatial correlation (see Fig.~\ref{fig:MLvsGANb}), the required number of photons per ray to achieve single-error-per-sample reduced from 5000 to the range between 320 and 400 for maximum-likelihood estimation. This is attributed to a more proper prior than the maximum-likelihood approach assumed, which leads to a better quality than the maximum-likelihood estimate. The generative models are slightly worse than the maximum-likelihood estimate at lower photon cases. Limited improvements are visible as the generative models need 200 to 256 photons per ray to achieve single-error-per-sample, reducing the required total number of photons for high-fidelity reconstruction. This is from the learned prior that provides a better classification cut-off, and the learned cut-off only improves the reconstruction quality when there is sufficient information for maximum-likelihood estimation. Fig.~\ref{fig:cointoss}(b) is the case with a biased coin toss that has $p=0.18521$. Compared with Fig.~\ref{fig:cointoss}(a), the required number of photons per ray for maximum-likelihood estimation to achieve single-error-per-sample is slightly reduced to the range between 200 and 256. With a lower probability of having copper, the 3D objects are now more sparse. Therefore, less attenuation from the copper material leads to effectively more photons captured by the detector pixel, improving the quality of the limited-angle measurements. We also observe a cross-over between maximum-likelihood estimation and deep generative reconstructions in Fig.~\ref{fig:cointoss} that was not present in the case with imaging object from CircuitFaker. The maximum-likelihood estimation has lower bit error rate in the very low photon regions for independent coin toss objects. This can be explained by our previous interpretation of the learned prior: since there is no spatial correlation of the object can be learned, a learned classification cut-off may only reduce the quality of a poor reconstruction in the very low photon conditions, especially when all the physical priors have already been considered. These results confirm that the deep learning approach benefits from the learned prior: when the assumed prior in the iterative algorithm is not well-suited for the reconstruction object (as for the case of circuit reconstruction), the generative models can drastically improve the reconstruction quality. On the other hand, when the prior distribution itself is simple, and the assumed prior matches the distribution (for the independent coin toss object), then the generative models may only provide a marginal improvement over the iterative algorithm. 


\section{Conclusions}
\label{conclusion}
In this work, we demonstrated the performance of the learned prior using PGAN for X-ray tomography. Our PGAN utilizes the maximum-likelihood estimate with known physics to compensate for the inherent ill-conditioning of the problem. This estimation is then processed by the learned prior from the trained deep generative model. Therefore, the PGAN reconstruction is generated by leveraging both known physics in the iterative algorithm and learned prior from deep learning. We developed an in-house model dubbed CircuitFaker to produce 3D circuits as the tomographic objects, and showed that the learned prior dramatically improves the reconstruction quality compared to maximum-likelihood estimate under the condition of limited observation angles and limited photon fluxes. We demonstrated that the learned prior is indeed from the CircuitFaker by reconstructing objects created with Bernoulli trials, and observing that there is limited improvement due to the generative model.

Compared with methods without physics or with less physics assisting in training, PGAN can also reduce the photon requirement with limited projection angles to achieve a given error rate, which suggests that further separation of the physical priors may improve the effectiveness of the learned prior. 
The advantages of using a physics-assisted learned prior in X-ray tomography may further enable nanoscale imaging with limited angles and fewer photons.

\begin{backmatter}
\bmsection{Funding and acknowledgments}
This research at MIT is supported by IARPA contract number FA8050-17-C-9113. G.B. also acknowledges support by Singapore's National Research Foundation through the Intra-Create grant NRF2019-THE002-0006 ``Retinal Analytics via Machine leaning aiding Physics (RAMP).''  Sandia National Laboratories is a multimission laboratory managed and operated by National Technology and Engineering Solutions of Sandia LLC (NTESS), a wholly owned subsidiary of Honeywell International Inc., for the U.S. Department of Energy's National Nuclear Security Administration (NNSA) under contract DE-NA0003525.
Research at NIST of B.K.A. and Z.H.L. is supported in part by the Office of the 
Director of National Intelligence (ODNI), Intelligence Advanced
Research Projects Activity (IARPA), via agreements D2019-1908080004 and D2019-1906200003. The authors acknowledge the MIT SuperCloud and Lincoln Laboratory Supercomputing Center for providing HPC resources that have contributed to the research results reported within this paper.

\bmsection{Disclosures}
Z.H.L. discloses U.S. Patents 6,389,101 
and 7,967,507 as well as US patent application 17/700,884 ``Efficient Method for Tomographic Reconstruction in the Presence of Fresnel Diffraction'' \cite{Levine2021}.
The authors declare that there are no other conflicts of interest related to this article.
This paper is
Sandia technical report SAND2022-7407.

\bmsection{Disclaimers}
The authors state that the opinions expressed herein are our own and do not represent the opinions of the funding agencies.
This paper describes objective technical results and analysis. Any subjective views or opinions that might be expressed in the paper do not necessarily represent the views of the U.S. Department of Energy or the United States Government.
Mention of commercial products does not imply endorsement by the authors' institutions.

\bmsection{Data availability statement}
The scripts to produce our deep learning models are publicly available at~\cite{Guo_Physics-assisted_Generative_Adversarial_2021}.

\end{backmatter}

\appendix
\begin{center}
    \textbf{\large Appendices}
\end{center}

\section{Details on the Maximum-likelihood computation}
\label{app:MLEinfo}


The projective reconstruction Fortran~95 code using maximum likelihood method used here has been presented recently as parts of a study on scatter corrections in tomography~\cite{Levine2019} and another on diffractive tomography~\cite{Levine2021}.  The key point is that the maximum likelihood objective function as formulated by Sauer and Bouman~\cite{sauer_bouman1993} is minimized using the version of the Broyden-Fletcher-Goldfarb-Shanno algorithm known as L-BFGS-B~\cite{fletcher2013practical}.

Each set of projection takes approximately 0.3~s to prepare using the code, then a further 1~s to make a Poisson sample of the mean intensities calculated by the projections.  The Poisson step is somewhat slow because it is done with interpreted Python.  Running solo, the iterative reconstruction takes approximately 7~s on a 3.7~GHz processor.  Although the code is implemented with MPI, this work was done with each case having a single processor.  Typically, 2-3 times as many cases as processors are computed at once to take advantage of hyperthreading which stretches the computing time per processor to 10~s, leading to about a {30\%} gain in throughput.
The solution is converged to approximately 12 decimal digits with about 35 iterations. The reconstructed solution is bounded between 0 and 2, whereas the void areas have a value of 0 and the wiring has a value of 1.

The memory required is dominated by four times the number of voxels, which is $4\times2048 \approx 8000$ plus the number of measurements, $32\times32\times8\approx8000$ for a total of about 16\,000 double words or 128~kilobytes.  This relatively light computational burden was repeated $(21+16)\times2000 = 74\,000$ times for Fig.~\ref{fig:expcircuit}, which was the bulk of the computation which was achieved by running for 2-3 days on a very few workstations.

\section{Network architecture}
\label{app:networkArch}
The general description of the network architecture is given in Section~\ref{deep networks}. The code to generate the networks is publicly available on GitHub~\cite{Guo_Physics-assisted_Generative_Adversarial_2021}. Here, we include more information about the network architecture for reproducibility.

\begin{figure}
    \centering
    \includegraphics[width=106mm]{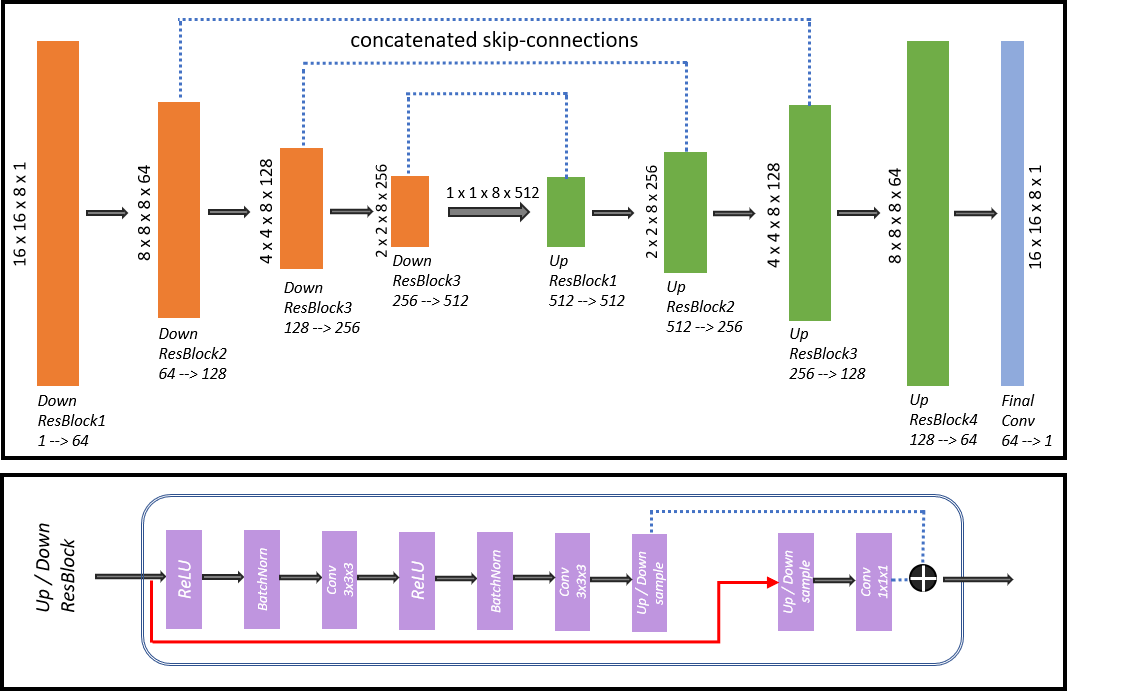}
    \caption{Network architecture for the deep generative model (generator).}
    \label{fig:arch}
\end{figure}

Fig.~\ref{fig:arch} is the detailed network architecture for the deep generative model (the generator). The overall design is based on UNet~\cite{ronneberger2015u} to perform pixel-by-pixel prediction (for 3D reconstruction, where the 3D object is voxelized by a 3D matrix). The input dimension to the model is in $(16, 16, 8, 1)$. Four DownResBlocks encode the input approximant and produce a latent representation that is in dimension of $(1, 1, 8, 512)$. Four UpResBlocks decode the latent representation to a vector in dimension of $(16, 16, 8, 64)$. Concatenated skip-connections are used in between the last three DownResBlocks and the first three UpResBlocks to preserve high frequency information of the input approximant~\cite{deng2020interplay}. Dropout layers are included to prevent over-fitting. The final layer of convolution reduces this vector to a final output in $(16, 16, 8, 1)$, and a Tanh layer forces the final output to the range between -1 and 1. The DownResBlock and UpResBlock share similar topology to the Resblock in ResNet~\cite{he2016deep}, except the use of different 3D sampling layers. Here, we implemented downsampling and upsampling layers that only sample the dimension in height and width but not depth. 

For the base generative model, feature extraction in the DownResBlock and UpResBlock is achieved by 3D convolutional kernel with spectral normalization~\cite{miyato2018spectral}. For the axial-attention based model, feature extraction in the DownResBlock is achieved by the mixture of 3D convolutional kernel and axial attention both with spectral normalization.

For models including the wavelet scattering transform, the wavelet representation for the input approximant is first produced by \textsc{HarmonicScattering3D} in the Kymatio package~\cite{JMLR:v21:19-047} with $J=2$ (maximum scale of $2^2$), integral powers with $\{0.5, 1.0, 2.0, 3.0\}$. Then, the batch normalization layers in the UpResBlock are replaced by conditional batch normalization (CBN) layers~\cite{de2017modulating, dumoulin2016learned, li2018adaptive}, where the conditional information is the wavelet representation. Note that the fully connected layers within the CBN are spectrally normalized as well.

The discriminator for all the generative models is the same, with four DownResBlocks bringing the input from dimension $(16, 16, 8)$ to $(2, 2, 1024)$, following with a reduce sum operation to bring it further to a vector of $(1, 1, 1024)$. A fully connected layer followed thereafter to produce a floating point number for classification.

For End-to-End methods, we add an additional DownResBlock at the top of the generator to down sample the measurements from $(32, 32, 8, 1)$ to $(16, 16, 8, 1)$. Concatenated skip-connections are also removed to avoid the issues of overlapping features from two different domains (from projections to object reconstruction).

\section{Network training}
\label{app:train}
Our proposed networks are implemented in Python 3.7.9 using TensorFlow 2.3.1, and trained with an NVIDIA V100 tensor core graphics processing unit on MIT Supercloud~\cite{reuther2018interactive}.  An Adam optimizer~\cite{kingma2014adam} is used with parameters $\beta_1 = 0.9$ and $\beta_2 = 0.999$. The two time-scale update rule (TTUR) is used to stabilize the training of the generative network~\cite{radford2015unsupervised, heusel2017gans}, where the initial learning rate is $10^{-4}$ for the generator and $4\times10^{-4}$ for the discriminator. In each iteration, the generator is updated four times while the discriminator is updated once.  

Training sets of 1800 reconstructions are generated independently for each condition studied, except that the ground truth is common. The batch size for training is 20. An additional 200 reconstructions per condition are used for testing. The learning rate is reduced by half when the validation loss stops improving for 5 iterations. We set the maximum number of iterations to be~200,
and the training stops early when either the validation loss plateaus for 20 iterations, or the minimum learning rate $10^{-8}$ is reached. This early-stop technique can prevent the model from over-fitting. The loss function for the autoencoder/generator consists of two parts: supervised loss and adversarial loss.  We choose supervised loss to be the negative of the Pearson correlation coefficient $r_{f, \Tilde{f}}$,
which is defined as 
\begin{equation}
r_{f, \Tilde{f}} = \frac{\text{cov}(f, \Tilde{f})}{\sigma_{f} \; \sigma_{\Tilde{f}}},
\end{equation}
where $\text{cov}$ is the covariance and $\sigma$ is the standard deviation.
The total objective of training is to find the optimal generator $G_{\text{opt}}$ given the approximant $\Tilde{f}$ and ground truth $f$:


\begin{equation}
   G_{\text{opt}}(\Tilde{f}) = \arg \min_{G} \max_{D} \mathbb{E}_{\left(f,\Tilde{f}\right)}\big\{-r_{f, G(\Tilde{f})} + \lambda \big[ \log D(f) + \log \big(1 - D(G(\Tilde{f}))\big) \big] \big\}.
\end{equation}
The hyper-parameter $\lambda$ controls the degree of generation from input noise to features. In our experiments, $\lambda$ ranges from $2^0$ to $2^{-6}$ with an incremental factor of 1/2.
The loss function for GAN is the hinge loss~\cite{lim2017geometric}, and is defined below:
\begin{equation}
\begin{split}
    L_D &= \text{mean} \big\{ \text{min} \{0, 1 - D(f)\} \big\} +  \text{mean} \big\{ \text{min} \{0, 1 + D(G(\hat{f}))\} \big\}, \\
    L_G &= - \text{mean} \{ D(G(\hat{f})) \}.
\end{split}
\end{equation}
Here, $L_G$ is the loss for generator and $L_D$ is the loss for discriminator. The operator min(...) chooses the smaller value between the two inputs. The mean is taken over the batch of the training data. 

\section{Convergence and stability of the deep generative network}
\label{app:convergenceAndStability}
As mentioned in Section~\ref{deep networks}, when first proposed, GANs had an instability problem during training. The model was easy to collapse, generating non-satisfactory results~\cite{cao2018recent}. It was since the appearance of deep convolutional generative adversarial networks (DCGAN)~\cite{radford2015unsupervised} that researchers began making GANs more stable by improving the structure and training skills. Later, the Wasserstein Generative Adversarial Network (WGAN) was introduced and provided a more detailed explanation of GANs' poor control~\cite{arjovsky2017wasserstein}. A solution was also proposed, i.e., imposing Lipschitz continuity, to improve the quality of generated results~\cite{arjovsky2017wasserstein, gulrajani2017improved}. Nowadays, there are well-known techniques to overcome the challenge in training GAN. We summarized the techniques we used in PGAN for interested researchers.
\subsection*{Spectral normalization}
While the WGAN approaches impose the Lipschitz continuity by gradient clipping or gradient penalty to stabilize the training, spectral normalization imposes a similar constraint by normalizing the weights within the network. This normalization technique is computationally light and easy to incorporate into existing implementations, and has been shown effective in many applications~\cite{brock2018large, lin2021spectral, Lan2020.06.09.143297}.

\subsection*{Hinge loss}
Hinge loss has shown improved performance when combined with spectral normalization.  Therefore, it has become standard in recent state of the art GANs ~\cite{kavalerov2021multi}.

\subsection*{Two time-scale update rule (TTUR)}
TTUR provides theoretical convergence of the GAN to a stationary local Nash equilibrium~\cite{heusel2017gans}. The core idea is to have an individual learning rate for both the discriminator and the generator. In our implementation, we choose $4\times10^{-4}$ for the discriminator and $10^{-4}$ for the generator.

\bibliography{ref}






\end{document}